\documentstyle[11pt,epsf]{article}

\newcommand{\del}{\partial}
\newcommand{\beq}{\begin{eqnarray}}
\newcommand{\eeq}{\end{eqnarray}}
\newcommand{\be}{\begin{eqnarray*}}
\newcommand{\ee}{\end{eqnarray*}}

\newcommand{\ra}{\rightarrow}

\newcommand{\ex}[1]{\langle\,#1\,\rangle}

\begin{document}

\centerline{\Large\bf {What is the Regularized Casimir Vacuum Energy Density?}}
\vskip 10mm
\centerline{Xinwei Kong \footnote{E-mail: kong@fys.uio.no} and Finn Ravndal \footnote{E-mail: finn.ravndal@fys.uio.no} }
\bigskip
\centerline{\it Institute of Physics}
\centerline{\it University of Oslo}
\centerline{\it N-0316 Oslo, Norway}

\bigskip
{\bf Abstract:} {\small The regularized total Casimir energy in spacetimes with boundaries is 
not in general equal to the integral of the regularized energy density. This paradoxical 
phenomenon is most transparently analyzed in the simple example of a massless scalar field
in 1+1 dimensions confined to a line element of length $L$ and obeying Dirichlet boundary
conditions.}



\vspace{10mm}
The quantum vacuum energy density due to zero-point fluctuations of a field in Minkowski space 
is in general a divergent quantity. In the absence of gravity it will not 
have any physical consequences and is thus usually ignored. However, if the quantum field is 
made to satisfy non-trivial boundary conditions, the vacuum energy density will still be
divergent, but in a different way from the previous case. The difference between these two results
is found to be finite and is called the Casimir vacuum energy density. The subtraction of two
divergent quantities to obtain a finite result is an example of a regularization method. 

Due to
the presence of boundaries one expects in general that the Casimir energy density is position 
dependent. On the other hand, in homogeneous spaces as on the surface of a sphere, the energy 
density must be constant because of symmetry. In either case, one would expect that the 
integrated  energy density equals the total Casimir energy, i.e. the regularized value of
the infinite sum
\beq
     E_0 = {1\over 2}\sum_n \omega_n                                          \label{E_0}
\eeq
in the case of non-interacting fields. Here $\omega_n$ are the eigenfrequencies of the wave 
equation for the given boundary conditions. We will in the following point out that this 
expectation does not in general hold. This was first realized by Deutsch and Candelas 
\cite{DC} who discovered that the field fluctuations and thus the vacuum energy density in
general diverges near confining boundaries.
 
Consider first a free Maxwell field. The vacuum energy density is then
\beq
     u = {1\over 2}(\ex{{\bf E}^2} + \ex{{\bf B}^2})                             \label{u}
\eeq
and is thus given by the sum $u = u_E + u_B$ due to fluctuations in the electric and magnetic
fields. The vacuum expectation values are most easily obtained in the simplest case of an 
electromagnetic field in a half-space bounded by an infinite plane. If ${\bf n}$ is the normal
to the plane in the $z$-direction, the standard, metallic boundary conditions are then
${\bf n}\times {\bf E} = {\bf n}\cdot {\bf B} = 0$. The regularized field fluctuations are 
found to be \cite{LR_1}
\beq
     \ex{{\bf E}^2} =  {3\over 16 \pi^2 z^4},        \hspace{10mm} 
     \ex{{\bf B}^2} = - {3\over 16 \pi^2 z^4}                             \label{plane}
\eeq
where $z$ denotes the distance from the plane. The Casimir energy density outside the plane
is thus exactly zero. 

The result (\ref{plane}) represent the leading terms in the vacuum fluctuations 
when you approach any bounding surface as shown by Deutsch and
Candelas \cite{DC}. In this limit the surface will be seen to be locally flat. For instance,
the vacuum energy density inside a spherical, metallic shell has been calculated by
Olaussen and Ravndal \cite{OR} and has this behaviour when one approaches the shell from the 
inside. Again the leading divergences cancel, but the sub-leading terms
do not and gives a divergent result for the energy density and thus also for the 
integrated Casimir 
energy. On the other hand, the total Casimir energy  obtained by Bender and Hays 
\cite{BH} from the infinite mode sum (\ref{E_0}) is finite.

While the calculation of the field fluctuations within a shell requires some numerical work, 
closed expressions have been obtained by L\"utken and Ravndal\cite{LR_2} for the generic 
Casimir system of two parallel metallic plates with separation $L$. The regularized results 
for the fluctuations at a distance $z$ from the left plane can be written as
\beq
     \ex{{\bf E}^2} = -{\pi^2\over 16L^4}\left({1\over 45} - F(\theta)\right), \hspace{10mm}
     \ex{{\bf B}^2} = -{\pi^2\over 16L^4}\left({1\over 45} + F(\theta)\right)   \label{plates}
\eeq
where $\theta = z\pi/L$ is the scaled distance and the function
\beq
     F(\theta) = - {1\over 2}{d^3 \over d\theta^3} \cot{\theta}  = {3\over\sin^4{\theta}} - {2\over\sin^2{\theta}}          \label{F}
\eeq
gives the position dependence of the fluctuations. In the limit $L \ra \infty$ we have essentially
only one plate and the result (\ref{plane}) is recovered. We see that although both the electric
part $u_E$ of the vacuum energy density and the magnetic part $u_B$ diverge near the plates, 
their sum $u = -\pi^2/720L^4$ is constant and just equal to the average of the total Casimir 
energy\cite{Casimir}. 

Virtual electrons in the vacuum affects the photons at low energies via the effective
Euler-Heisenberg interaction\cite{EH}
\beq
    {\cal L}_{EH} = {2\alpha^2\over 45 m^4}\left[({\bf E}^2 - {\bf B}^2)^2 
                  + 7 ({\bf E}\cdot{\bf B})^2\right]                           \label{EH}
\eeq
where $\alpha$ is the fine structure constant and $m$ the elctron mass. Its effect on the
propagation of light between two parallel metallic plates have been investigated by 
Scharnhorst\cite{Scharnhorst} and Barton\cite{Barton}. It will also give a correction to
the above free result for the vacuum energy density\cite{KR}. Then there is no 
longer any exact cancellation between the divergences in the electric and magnetic contributions
near the plates so that the regulated energy density will diverges there. When integrating it 
over the volume between the plates one thus gets a divergent result. This is physically 
meaningless and similar to what was found for free photons confined within a spherical 
shell\cite{OR}.

In order to analyze this problem in a more transparent way, let us instead simplify it
by ignoring the degrees of freedom parallell to the walls. Effectively, we can then instead
consider a massless scalar field $\phi(t,z)$ confined on the $z$-axis between $z=0$ and $z=L$.
In the free Lagrangian  ${\cal L} = (1/2)[(\del_t\phi)^2 - (\del_z\phi)^2]$ we can consider
$\del_t\phi$ to be the electric field and $\del_z\phi$ to be the magnetic field. We can thus 
continue to talk about the electric $u_E = (1/2)\ex{(\del_t\phi)^2}$ and magnetic $u_B = 
(1/2)\ex{(\del_t\phi)^2}$ contributions to the vacuum energy density.
Imposing the Dirichlet condition that the field vanishes at the boundaries, the normalized
eigenmodes of the field are
\beq
       \phi_n(z) = \sqrt{2\over L}\sin{(\omega_n z)}     \label{modes}
\eeq
where $\omega_n = \pi n/L$ with $n= 1,2,3,\ldots{}$. The total vacuum energy (\ref{E_0}) is then
\beq
     E_0 = {\pi\over 2L}\sum_{n=1}^\infty n = -{\pi\over 24 L}            \label{E0}
\eeq
when we use zeta-function regularization where $\zeta(-1) = -1/12$. Assuming a constant vacuum
energy density, it is therefore $u = -\pi/24 L^2$.

Let us now consider the partial contributions to the vacuum energy density. The part coming from
fluctuations in the electric field is given by the divergent sum
\beq
       u_E &=& {1\over 2L}\sum_{n=1}^\infty \omega_n \sin^2{(\omega_n z)} 
            = {1\over 4L}\sum_{n=1}^\infty \omega_n (1 - \cos{2\omega_n z})       \label{uE}
\eeq
Again using zeta-function regularization, the first sum is given by $\zeta(-1)= -1/12$ while
the second sum becomes
\beq
       \sum_{n=1}^\infty n \cos(2\theta n) = {1\over 4}{d\over d\theta} \cot{\theta}
                                           = -{1\over 4\sin^2\theta}                \label{sum}
\eeq 
These finite results can also be obtained by using other regularization schemes, e.g.  a 
high-frequency cutoff or splitting the positions of the two fields in the correlator. The
regulated result for this part of the vacuum energy is therefore 
\beq
       u_E = -{\pi\over 16L^2}\left({1\over 3} - {1\over \sin^2\theta}\right)   \label{u_E}
\eeq
The constant first term is just one half of the mean energy density $u = -\pi/24L^2$. The other
half will be provided by a similar term in the magnetic contribution. All the position dependence
lies in the second term which is positive definite and diverges near the endpoints of the 
confining region. When it is integrated up it gives an infinite contribution. That is not a
real problem yet because there will be a corresponding infinite contribution from the magnetic
part which will come with an opposite sign so that they cancel. But the seemingly paradoxical
situation arises when we integrate the electric energy density on the form (\ref{uE}) before 
regularization. Then the each mode in the second, infinite sum gives zero because of the boundary
conditions,
\beq
     \int_0^L\!dz\, u_E = {1\over 4L}\sum_{n=1}^\infty  \omega_n
     \left(L - {1\over 2\omega_n}\sin{(2\omega_n L)}\right)  = -{\pi\over 48L}      \label{int}
\eeq   
So the position-dependent part of the vacuum energy density contributes 
nothing to the total Casimir energy. This is a concrete example of the realization by Deutsch 
and Candelas\cite{DC} that the operations of integration and regularization do not commute in
general. The same obviously also holds for the magnetic energy density
\beq
     u_B = {1\over 2L}\sum_{n=1}^\infty \omega_n \cos^2{(\omega_n z)} 
            = -{\pi\over 16L^2}\left({1\over 3} + {1\over \sin^2\theta}\right)    \label{uB}     
\eeq
which diverges with opposite sign when approaching the endpoints of the interval. 

The structure of these two contributions to the vaccum energy density is exactly the same as 
for the full Maxwell field in (\ref{plates}). Only the first, position-independent part will
contribute to the total Casimir energy for this configuration of two parallel plates. On the
other hand, for a spherical shell the position-dependent terms in $u_E$ and $u_B$ don't cancel
\cite{OR} and it is therefore not completely clear what the physical significance should be
given to the resulting total energy density which now diverges when one approaches the shell.
The corresponding correlators $\ex{{\bf E}^2}$ and $\ex{{\bf B}^2}$ can in principle be
measured. For instance, they will affect the energy levels of an atom placed in the vicinity of 
such a metallic boundary \cite{LR_2}\cite{QED}. By moving the atom around one can then map the 
spatial variation of the vacuum fluctuations by spectroscopic methods.

Scharnhorst\cite{Scharnhorst} and Barton\cite{Barton} could show that by extending the Maxwell 
theory of light by including the non-linear Euler-Heisenberg interaction (\ref{EH}), 
the refractive index is changed in the vacuum between two plates. This is again due to the 
non-zero values of $\ex{{\bf E}^2}$ and $\ex{{\bf B}^2}$ in (\ref{plates}). Somewhat 
surprisingly, the position dependence in both of these correlators cancel out in the net result 
for the refractive index. Again it is only the constant terms which contribute to the physics.

One can emulate the Euler-Heisenberg interaction in our one-dimensional system by considering the
scalar Lagarangian
\beq
     {\cal L} = {1\over 2}(\del_\mu\phi)^2 + {\alpha\over m^2}(\del_\mu\phi)^4    \label{LEH}
\eeq
Here $\alpha$ is some small dimensionless constant and $m$ a heavy mass. The Lagrangian is 
invariant under the field transformation $\phi \ra \phi + const$ and thus describes massless
particles. Treating the interaction
in lowest order perturbation theory, one easily finds the resulting vacuum energy density 
after regularization to be
\beq
    u = - {\pi\over 24L^2} -  {\alpha\pi^2\over 8m^2L^4}\left({1\over 18} 
        + {1\over\sin^4\theta}\right)                                            \label{uu}
\eeq
Now it diverges near the endpoints where $\theta \ra 0$ and the integrated Casimir energy is
infinite. However, if we integrate the energy density and then regularize as we did in 
(\ref{int}), we get the finite result
\beq
    E_0 = - {\pi\over 24L} -  {\alpha\pi^2\over 144m^2L^3}                        \label{E1}
\eeq
The potentially divergent part vanishes now vansihes since  $\zeta(-2) = 0$. We see that the 
total Casimir 
energy is furnished by just the constant part of energy density (\ref{uu}). Again there is no
contribution from the position-dependent terms.

We have similarly calculated the total Casimir energy for the photons  between two plates
interacting via the Euler-Heisenberg interaction\cite{KR}. The regulated energy density 
diverges near the walls while the integrated energy is again finite after regularization 
and equals
\beq
    E_0 = - {\pi^2\over 720L^3} 
          - {11\alpha^2\pi^4\over 2^7  3^5  5^3 m^4L^7}               \label{E3}
\eeq
This new quantum correction to the standard Casimir effect is obviously much too small to be 
detected by present methods.

\end{document}